\documentstyle[preprint,tighten,aps,epsfig,colordvi,psfig]{revtex}

\def\spose#1{\hbox to 0pt{#1\hss}}
\def\ltapprox{\mathrel{\spose{\lower 3pt\hbox{$\mathchar"218$}}
 \raise 2.0pt\hbox{$\mathchar"13C$}}}
\begin{document}
\draft
\vskip 2cm

\title{Free Energy and Plaquette expectation value for gluons on the
  lattice, in three dimensions}

\author{H. Panagopoulos$^{a}$, A. Skouroupathis$^{a}$ and A. Tsapalis$^{b}$}
\address{$^{a}$Department of Physics, University of Cyprus, P.O. Box 20537,
Nicosia CY-1678, Cyprus \\
{\it email: }{\tt haris@ucy.ac.cy, php4as01@ucy.ac.cy}\\
$^{b}$University of Athens, Institute of Accelerating Systems and Applications, Athens, Greece \\
{\it email: }{\tt a.tsapalis@iasa.gr}}
\vskip 3mm

\date{\today}

\maketitle

\begin{abstract}

We calculate the perturbative value of the Free Energy in 
Lattice QCD in three dimensions, up to three loops. Our 
calculation is performed using the Wilson formulation for gluons in
$SU(N)$ gauge theories. 

The Free Energy is directly related to the average plaquette.
To carry out the calculation, we compute the coefficients 
involved in the perturbative expansion of the Free Energy
up to three loops, using an automated set of procedures developed by us in
Mathematica. The dependence on $N$ is shown explicitly in our
results. 

For purposes of comparison, we also present the individual
contributions from every diagram. These have been obtained by means of
two independent calculations, in order to cross check
our results.

\medskip
{\bf Keywords:} 
Lattice QCD, Lattice perturbation theory, Free Energy.

\end{abstract}

\newpage


\section{Introduction}

In this paper we calculate the Free Energy of QCD on the lattice, 
up to three loops in perturbation theory. We are 
interested in the three dimensional case, involving only
gluons.

The choice of studying the Free Energy is motivated by the fact that,
apart from being a simple physical observable which can be calculated
in perturbation  
theory, it is used to determine the static thermodynamic properties 
of a physical system. Furthermore, this quantity is ideal for studying various 
aspects of QCD. For example, the Free Energy is often employed in the
context of thermal QCD, for the purpose of characterizing the
deconfinement transition 
between the low temperature regime, ruled by confinement, and the high 
temperature regime ruled by asymptotic freedom. Asymptotic 
freedom guarantees a small coupling constant at large temperatures,
which in turn allows for a perturbative evaluation of observables. In
addition to that, 
the perturbation expansion of the Free Energy is involved in the 
determination of the ``finite part'' of the gluon condensate in lattice 
regularization. 

Finally, the plaquette expectation 
value, which is directly related to the Free Energy, is considered in 
the study of an effective 3-D pure gauge theory called Magnetostatic QCD
(MQCD)~\cite{Hietanen}, which is matched to QCD using perturbation
theory. Due to the superrenormalizable nature of the 3-D theory, only a finite
number of divergences arise and these can be computed perturbatively,
allowing for a complete match between the lattice regularization scheme and a
continuum scheme such as ${\overline {MS}}$.
MQCD is an instance of ``dimensional reduction'', whereby the static 
properties of a (3+1)-dimensional field theory at high temperature can be 
expressed in terms of an effective field theory in 3 space
dimensions. Dimensional reduction has been applied to QCD (see,
e.g.,~\cite{Braaten1,Braaten2}), in order to 
resolve a longstanding problem  
regarding the breakdown of the perturbation expansion for the Free Energy~\cite{Linde}.

The evaluation of the Free Energy in four dimensions has already been carried 
out to 2 loops~\cite{GR} and 3 loops~\cite{ACFP,Panago}. In three
dimensions, the only results available so far have been obtained
through the method of stochastic perturbation theory~\cite{DiRenzo}.
Our results can be used both for comparison with existing results extracted
from different methods, or for calculations involving the perturbative 
expansion of the Free Energy. With respect to this, the coefficients of the 
perturbative expansion allow the subtraction of all the divergent 
contributions from the gluon condensate~\cite{Kajantie} and improve the
previous estimates of Ref.~\cite{DiRenzo}. 

\section{Calculation}

In standard notation the Wilson action for gluons reads:

\begin{equation}
S_L =  {1\over g_0^2} \sum_{x,\,\mu,\,\nu}
{\rm Tr}\left[ 1 - U_{\mu,\,\nu}(x) \right]  
\label{glact}
\end{equation}

Here $U_{\mu,\,\nu}(x)$ is the usual product of link variables
$U_{\mu}(x)$ along the perimeter of a plaquette in the $\mu$-$\nu$
directions, originating at $x$;
$g_0$ denotes the bare coupling constant.
Powers of the lattice spacing $a$ have been omitted and 
may be directly reinserted by dimensional counting.

We use the standard covariant gauge-fixing term~\cite{KW}; in terms of
the gauge field $Q_\mu(x)$ $\left[U_{\mu}(x)=
  \exp(i\,g_0\,Q_\mu(x))\right]$, it reads:
\begin{equation}
S_{gf} = \lambda_0 \sum_{x,\mu , \nu}
\hbox{Tr} \, \bigl\{ \Delta^-_{\mu} Q_{\mu}(x) \Delta^-_{\nu}
Q_{\nu}(x)\bigr\} , \qquad
\Delta^-_{\mu} Q_{\nu}(x) \equiv Q_{\nu}(x - {\hat \mu}) - Q_{\nu}(x).
\end{equation}

Having to compute a gauge invariant quantity, we chose to work
in the Feynman gauge, $\lambda_0 = 1$. 
Covariant gauge fixing produces the following
action for the ghost fields $\omega$ and $\overline\omega$

\begin{eqnarray}
&\displaystyle S_{gh} = 2 \sum_{x,\mu} \hbox{Tr} \,
\biggl\{ \Bigl(\Delta^+_{\mu}\omega(x)\Bigr)^{\dagger} \Bigl( &\Delta^+_{\mu}\omega(x) +
i g_0 \left[Q_{\mu}(x),
\omega(x)\right] + \frac{i\,g_0}{2}
\,\left[Q_{\mu}(x), \Delta^+_{\mu}\omega(x) \right] \nonumber\\
 & & - \frac{g_0^2}{12}
\,\left[Q_{\mu}(x), \left[ Q_{\mu}(x),
\Delta^+_{\mu}\omega(x)\right]\right]\nonumber\\
&&  - \frac{g_0^4}{720}
\,\left[Q_{\mu}(x), \left[Q_{\mu}(x), \left[Q_{\mu}(x), \left[ Q_{\mu}(x),
\Delta^+_{\mu}\omega(x)\right]\right]\right]\right]
+ \cdots \Bigr)\biggr\} ,\nonumber\\
&\Delta^+_{\mu}\omega(x) \equiv \omega(x + {\hat \mu}) - \omega(x).&
\end{eqnarray}

Finally the change of integration variables from links to vector
fields yields a Jacobian that can be rewritten as 
the usual measure term $S_m$ in the action:

\begin{equation}
S_m = \sum_{x,\mu} \biggl\{ \frac{N g_0^2}{12}\, \hbox{Tr} \,
\bigl\{ \bigl(Q_\mu(x)\bigr)^2 \bigr\} + \frac{N g_0^4}{1440} \,\hbox{Tr} \,
\bigl\{ \bigl(Q_\mu(x)\bigr)^4 \bigr\} + \frac{g_0^4}{480} \,\Bigl(\hbox{Tr} \,
\bigl\{ \bigl(Q_\mu(x)\bigr)^2 \bigr\}\Bigr)^2 + \cdots \biggr\}
\end{equation}

In $S_{gh}$ and $S_m$ we have written out only
terms relevant to our computation.
The full action is: 

\begin{equation}
S = S_L + S_{gf} + S_{gh} + S_m.
\end{equation}

In D-dimensions, the average value of the action density, $S/V$, is
directly related to the average $1\times 1$ plaquette $P$: 
\begin{equation}
\langle S/V \rangle = {D\,(D-1)\over 2} \,\beta\,\langle E(P)\rangle.
\end{equation}
where:
 \begin{equation}
E(P)= 1 - {1\over{N}}\,{\rm Re}[{\rm Tr}(P)],\qquad \beta = {2\,N\over g_0^2}
\label{E(P)}
\end{equation}

We will calculate $\langle E\rangle$ in perturbation theory:
\begin{equation}
\langle E \rangle = c_1 \; g_0^2 + c_2 \; g_0^4 + c_3 \; g_0^6 + \cdots
\label{expansion}
\end{equation}

The $n$-loop coefficients $c_n$ have been known for some time up to 3 
loops in the 4-dimensional theory~\cite{ACFP}. 

The calculation of $c_n$ proceeds most conveniently by computing first the Free Energy $-(\ln Z)/V$, where $Z$ is the full partition function:
\begin{equation}
Z \equiv \int [{\cal D}U] \exp(-S) .
\label{Z}
\end{equation}
The average of $E$ is then extracted as follows:
\begin{equation}
\langle E \rangle = - \biggl({D\,(D-1) \over 2}\biggr)^{-1}\, {\partial
  \over {\partial \beta}}\, \left( {\ln Z \over V} \right) . 
\label{E}
\end{equation}
In particular, the perturbative expansion of $(\ln Z)/V$ is:
\begin{equation}
(\ln Z)/V = d_0 - {(D-1)\over 2}\,(N^2{-}1)\,\ln\beta + {d_1\over\beta} + {d_2\over\beta^2} + \cdots
\label{lnZ/V}
\end{equation}
We can now write the perturbative expansion of $\langle E \rangle$, by
combining Eqs. (\ref{E}) and (\ref{lnZ/V}):
\begin{eqnarray}
\langle E \rangle = \biggl({D\,(D-1) \over 2}\biggr)^{-1}\,\Biggl[{(D-1)\over 2}\,{(N^2{-}1)\over\beta}+{d_1\over\beta^2}+{2d_2\over\beta^3}+ \cdots\Biggr]
\label{Eexpansion}
\end{eqnarray} 
The coefficients $c_n$ involved in Eq. (\ref{expansion}) can be specified by the general expression 
\begin{equation}
c_n = \biggl({D\,(D-1) \over 2}\biggr)^{-1}\,{(n-1)\,d_{n-1}\over
  (2N)^n}\,, \qquad n\ge 2
\label{c_n}
\end{equation}
which leads immediately to the following relations for $D=3$:
\begin{eqnarray}
c_2 &=& d_1/(12N^2), \\ 
c_3 &=& d_2/(12N^3).
\label{c2c3}
\end{eqnarray}

\section{Results}

A Total of 36 Feynman diagrams contribute to the present calculation,
 up to three loops, as shown in Figure 1. The involved algebra of the lattice 
perturbation theory was carried out using our computer
package in Mathematica. The value for each diagram is computed
numerically for a sequence of finite lattice sizes $L$, followed by an
extrapolation to $L\to\infty$ (see next Section). The finite-L results, on a
per diagram basis, are available upon request from the authors. We performed 
two independent calculations for each diagram, in order to verify our results.

Certain diagrams, considered individually, are infrared divergent;
they contain subdiagrams which renormalize the gluon propagator to one
loop, as shown in Figure 2. Taken as a group, these diagrams become
infrared finite and their value can be extrapolated to infinite lattice size;
extrapolation leads to a (small) systematic error, which 
is estimated quite accurately. 

The coefficients $c_i$ involved in the perturbative expansion of $\langle E
\rangle$ can be calculated by summing all contributing diagrams. They
take the form:
\begin{eqnarray}
 c_1 &=& {{N^2 {-} 1} \over 6N}\, , \nonumber\\
 c_2 &=& \sum_{j}{(N^2-1)\over 6N} \left( {c_{2,0}^{(j)}\over N} + c_{2,1}^{(j)}N \right), \\
 c_3 &=& \sum_{j}{(N^2-1)\over 3N} \left( {c_{3,0}^{(j)}\over N^2} + c_{3,1}^{(j)} +{c_{3,2}^{(j)}\,N^2} \right) . \nonumber
\end{eqnarray}

\begin{center}
\psfig{figure=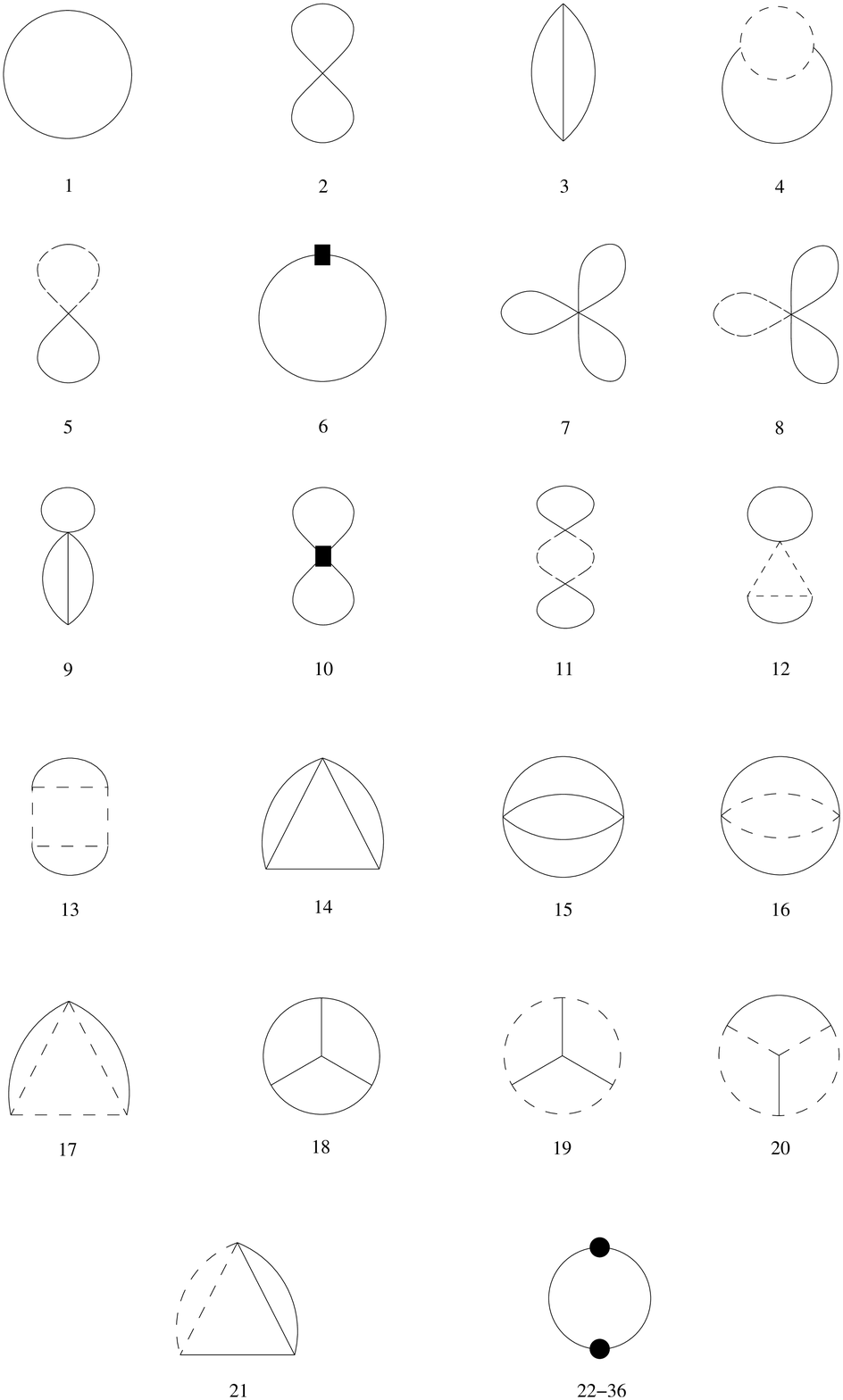,scale=0.55}\\
\medskip
{\footnotesize Fig. 1 Diagrams contributing
  to the Free Energy at one loop (1), two loops (2-6) and three loops
  (7-36). Solid (dashed) lines represent gluons (ghosts). The 
  filled square is the contribution from the ``measure'' part of the
  action.
 Black bullets stand for an effective vertex, which is shown
  in Fig. 2.}
\end{center}
\vfill
\eject

\begin{center}
\psfig{figure=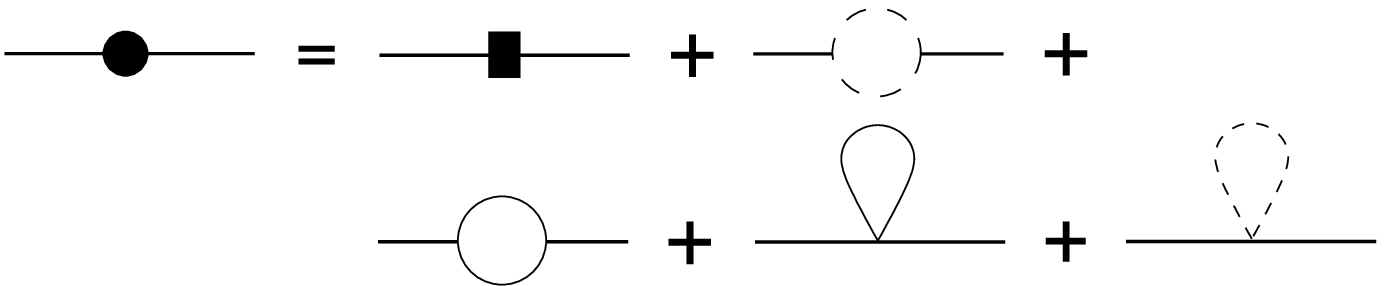,scale=0.8}\\
\medskip
{\footnotesize Fig. 2 Effective vertex involved in diagrams 22-36. The
  filled square is the contribution from the ``measure'' part of the
  action. }
\end{center}

The index $j$ runs over all contributing diagrams. The coefficients
$c_{2,k}^{(j)},\, c_{3,k}^{(j)}$ are pure numbers and we calculate
them for each diagram separately. Their numerical values are listed
in Tables \ref{c2table} and \ref{c3table}. For some diagrams, the coefficients 
are known analytically. The expressions are presented as a function of the 
1-loop vacuum diagram $P_1$ (see below):
\begin{eqnarray}
c_{2,1}^{(2)}&=&(1+ 28 P_1 - 72 P_1^2)/48 \nonumber \\
c_{2,1}^{(5)}&=&-P_1/12 \nonumber \\
c_{2,1}^{(6)}&=&-P_1/8  \nonumber \\
c_{3,1}^{(7)}&=&(504 P_1 -1296 P_1^2 -6)/5184 \nonumber \\
c_{3,2}^{(7)}&=&(3240 P_1^3 -792 P_1^2 - 63 P_1 -1)/5184 \nonumber \\
c_{3,2}^{(8)}&=&-P_1^2/288 \nonumber \\
c_{3,2}^{(10)}&=&-P_1^2/384 \nonumber \\
c_{3,2}^{(11)}&=&-P_1^2/288 \nonumber \\
\end{eqnarray}
The value of $P_1$ can be expressed in terms of the Bessel function of
the first kind $I_0$:
\begin{eqnarray}
P_1&=&\int_{-\pi}^{\pi} {d^3p \over (2\pi)^3}{1 \over 4{\displaystyle
    \sum_{\mu=1}^{3}}{\rm
    sin}^2(p_{\mu}/2)}=\int_{0}^{\infty}dt\bigg[\int_{-\pi}^{\pi}{dp
    \over (2\pi)}\, e^{{\displaystyle -t \,4\,{\rm
	sin}^2(p/2)}}\bigg]^3=\int_{0}^{\infty}dt\bigg[e^{-2t}I_0(2t)\bigg]^3 \nonumber \\ 
&=&0.2527310098590\cdots
\label{P1}
\end{eqnarray}
The ghost loop in diagram 21 leads to a vanishing contribution by color 
antisymmetry.

Using our results we can express the average Free Energy as a function of 
$g_0$, for specific values of $N$:
\begin{equation}
N=2:\quad \quad \langle E
\rangle=(1/4)\,g_0^2+0.01453916(1)\,g_0^4+0.0053459(2)\,g_0^6 + \cdots
\label{EgN2}
\end{equation}
\begin{equation}
N=3:\quad \quad \langle E
\rangle=(4/9)\,g_0^2+0.05420318(2)\,g_0^4+0.0317648(7)\,g_0^6 + \cdots
\label{EgN3}
\end{equation}

Presented in Fig. 3 is the behaviour of the average Free Energy as a function of $1/\beta$, for $N=2$ and $N=3$. 

\newpage
\phantom{a}
\vskip 6cm
\begin{center}
\input{./graph1.pslatex}
\end{center}

\newpage
\section{Discussion and Conclusions}

Several remarks are in order regarding our computation. 

\noindent
$\bullet$\ \ As mentioned
above, certain subsets of diagrams are infrared finite only when
summed together, and they correspond to insertions of the one-loop
renormalized gluon propagator. Similarly, diagrams 11, 12 and 13
renormalize the ghost propagator; however, they are separately finite. 

\noindent
$\bullet$\ \ The manipulation of the algebraic expressions in order to
reach a code for 
numerical loop integration, is completely done in Mathematica. The 
production of vertices as 
well as their contraction is fully automated. A number of routines 
render automatic the simplification of the expressions (such as color 
contractions and Lorentz index manipulation) and
exploitation of their symmetries. We have also developed an efficient 
three dimensional ``integrator'' to automatically convert the integrand 
to Fortran code. The size of a typical input integrand is up to 
$\ltapprox 1000$ terms after 
simplification. Finally, the extrapolation of data to 
infinite $L$ is also automated. The only part which is done by hand is
the enumeration of diagrams, but this is simple enough to do, even up 
to 4 loops. Each $n$-vertex diagram is represented by an $n\times n$
``incidence'' matrix $M$, where $M_{ij}$ is the number of gluons
joining vertices $i$ and $j$\, and by a similar matrix for ghosts.
The procedure described above is not CPU 
intensive as regards symbolic manipulations in Mathematica. 
Of course, this is due 
to the usage of Wilson action for gluons (improved actions are more 
cumbersome in terms of CPU as well as RAM).       

\noindent
$\bullet$\ \ Numerical integrals are performed for finite lattice
sizes\footnote{A detailed  
treatment of the zero modes on a finite lattice is not necessary for the
purposes of extrapolation to $L\to\infty$}, up to $L=100^3$, and
the results are subsequently extrapolated to $L\to\infty$. Diagrams
18-20 (those having the form of the Mercedes emblem) are more CPU
intensive and require approximately 40 hours each (for $L=40^3$) on a recent 
PentiumIV computer (3.2 GHz); for these diagrams, integration over 
the three loop momenta can only be performed in a nested fashion, rather 
than in parallel. Fortunately, lattice sizes up to $L=40^3$ already lead to a
sufficiently small systematic error in these cases. For the rest of the 
diagrams, two out of the three loop momenta can be integrated in 
a parallel fashion, and the calculation requires approximately 
1 hour per diagram for $L=100^3$. 

Extrapolation to infinite lattice size is of course a source of
systematic error. To estimate this error, our procedure
carries out the following steps: First, different
extrapolations of our numerical data $r_L$ are performed using a broad
spectrum of functional 
forms $f^k(L)$ (around 50 of them) of the type:
\begin{equation}
f^k(L) = \sum_{i,j} e^{(k)}_{i,j}\, L^{-i}\, (\ln L)^j
\label{extrapolationFit}
\end{equation}

A total of $N_k$ coefficients appear in the $k^{\rm th}$ functional
form; these coefficients are determined uniquely using the results on
$N_{k}$ lattices of consecutive size $L$. 

For the $k^{\rm th}$ such extrapolation, a deviation $d_k$ is
calculated using alternative criteria for quality of fit. 
One possible criterion, as an example, is the difference:
$d_k = \bigl|f^k(L^{^*}) - r_{L^{^*}}\bigr|$, where
$L^{^*}$ is the largest lattice size which was not used
in the determination of $e^{(k)}_{i,j}$\,.
Finally, these
deviations are used to assign weights $d_k^{-2}/(\sum_k d_k^{-2})$ to
each extrapolation, producing a final value together with
the error estimate. 
We can check the reliability of our error estimates in a number of
ways: In those cases where the result for a diagram is also known
analytically, this result coincides with the numerical one within the
systematic error; also, extrapolations which incorporate new data from
larger lattices are compatible with previous results, again within
systematic error.

The 3-dimensional case is slightly different and more delicate than the 
case of 4 dimensions; in the latter only even powers of $1/L$ must be used
($i={\rm even}$ in Eq.(\ref{extrapolationFit})), along with possible 
logarithms. In 3 dimensions one must also use odd 
negative powers of $L$, since the main 
difference between a sum over discrete values of momenta and the
corresponding integral comes from omitting the contribution of the
zero mode. This quantity behaves like $I=\int_{_V}d^3p/p^2$ where $V$
is a sphere of diameter $2\pi/L$\,. We see
that $I=4\pi^2/L$. 

Using larger lattices is of course computationally more feasible in
3 dimensions. At the same time, it is also more necessary than in 4-D, since
the infrared behaviour of loop integrals worsens in 3-D. Suffice it to
recall that 4-loop integrals in three dimensions will bring about a
genuine divergence, requiring introduction of an IR regulator.

\noindent
$\bullet$\ \ Referring to diagrams 22-36, integration in 3-D with use of an IR
regulator (in our case, $L$) may give extra terms
which do not exist in the 4-D case, leading to potentially wrong results, when
the regulator is taken to its limit value ($L\rightarrow \infty$). Due
to the presence of a loop
involving two propagators at the same momentum, each of these diagrams
exhibits a behaviour of the type: 

\begin{equation}
(a + {b \over L})(cL + d)
\label{spurious}
\end{equation}

After summing all diagrams and extrapolating to the limit
$L\rightarrow \infty$, the result we get has the form $ad + bc$
(terms like $acL$ add up to zero). Of course, the terms $bc$ are
spurious and must not be taken into account. A similar procedure in 4
dimensions leads instead to:

\begin{equation}
(a + {b \over L^2})(c \, \ln L + d)
\end{equation}
so that spurious terms are absent: $(b/L^2)(c\,\ln L)\to 0$.
To solve this problem, we can write the mathematical expression for
each vertex involved in diagrams 22-36 (Fig. 2) in a subtracted way:
$f(p)=f(0)+(f(p)-f(0))$, where $p$ is the momentum on external
lines. Once all the subtracted vertices are summed, 
the terms $f(0)$ cancel among themselves. In addition to that, the terms
$f(p)-f(0)$ are at least of first order in $p$ and factors of the type
$cL$ involved in Eq. (\ref{spurious}) cannot arise. The subtracted
diagrams take the form $(\tilde{a}+ \tilde{b} / 
L)(\tilde{d})$ and thus, extrapolation gives the desired results
$\tilde{a}\tilde{d}$. 

\noindent
$\bullet$\ \ For purposes of comparison, we can express the perturbative expansion of the 
Free Energy shown in Eqs. (\ref{EgN2}) and (\ref{EgN3}), as a function
of $1/\beta$: 

\begin{eqnarray}
N=2:&\quad \quad &\langle E \rangle=\bigg({1 \over
  \beta}\bigg)+0.2326265(2)\,\bigg({1 \over
  \beta}\bigg)^2+0.34214(1)\,\bigg({1 \over \beta}\bigg)^3 + \cdots
\label{EbN2}\\
N=3:&\quad \quad &\langle E \rangle=(8/3)\,\bigg({1 \over
  \beta}\bigg)+1.951315(1)\,\bigg({1 \over
  \beta}\bigg)^2+6.8612(2)\,\bigg({1 \over \beta}\bigg)^3 + \cdots
\label{EbN3}
\end{eqnarray} 

For the case $N=3$, the value $(2N)^3 c_3 = 6.8612(2)$ improves a
previous estimate 
$(2N)^3 c_3=6.90_{-0.12}^{+0.02}$\ ~\cite{DiRenzo}, which was obtained
with the  
method of stochastic perturbation theory; it allows for a more stable
extraction of the ``finite'' part of the gluon condensate, as expressed in
Ref.~\cite{Kajantie} (using the notation of that reference for the
symbols $c_i$):
\begin{equation}
8\,{(N^2-1)\,N^6\over (4\pi)^4}\,B_G = 
   \lim_{\beta\to\infty} \beta^4 \biggl\{\big\langle 1 - {1\over N}\, {\rm
   Tr} (P)\big\rangle - \Bigl[{c_1\over\beta} + {c_2\over\beta^2} + 
{c_3\over\beta^3} + {c_4\over\beta^4}\,\big(\ln \beta +
   c_4^\prime\big)\Bigr]\biggr\} 
\end{equation}
where the logarithmic coefficient $c_4$ is known from the continuum
vacuum energy density in the ${\overline {MS}}$ scheme:
\begin{equation}
c_4 = 8\, {(N^2-1)\,N^6\over (4\,\pi)^4}\,\Bigl({43\over 12} - 
{157\over 768}\,\pi^2\Bigr),\qquad
c_4(N{=}3) = 2.92942132...
\end{equation}
and $c_4^\prime$ requires a 4-loop calculation in lattice perturbation
theory. 

\noindent
$\bullet$\ \ A possible extension of the present calculation to 4
loops is feasible with the techniques used in this paper, but one  
must take into account some additional difficulties: First of all there 
is a total of 250 diagrams to be calculated  (43 pure gluon diagrams, 
115 diagrams with one ghost loop, 37 diagrams with two ghost loops, 4 diagrams
with three ghost loops and 51 diagrams with measure terms) and computation 
time is proportional to $L^{12}$ (in most diagrams, loop integrals are 
factorizable, with the exception of a ``nonplanar'' diagram and its two ghost 
variants). Furthermore, due to the existence of a genuine logarithmic 
divergence, one must insert an IR cutoff $m$ leading to results of the form 
$c\,{\rm ln}(m)+d$. Of course, the vertices and the contracted expressions, 
as well as a 4-loop version of our integrator, can still be produced
automatically in this case. 

\bigskip\noindent
{\bf Acknowledgements: } This work is supported in part by the
Research Promotion Foundation of Cyprus (Proposal Nr: $\rm ENTA\Xi$/0504/11).

\begin{table}[p]
\begin{center}
\begin{minipage}{8cm}
\caption{Per-diagram contributions to $c_2$.
\label{c2table}}
\begin{tabular}{cr@{}lr@{}l}
\multicolumn{1}{c}{Diagram}&
\multicolumn{2}{c}{$c_{2,0}$} &
\multicolumn{2}{c}{$c_{2,1}$} \\
\tableline \hline
2	&-1/&12                       &0&.0724503107... \\
3	&0&	                      &0&.03394382(1)\\
4       &0&                           &-0&.00383019(1)\\
5       &0&                           &-0&.0210609174...\\
6       &0&                           &-0&.0315913762...\\
\hline
Sum & -1/&12                          &0&.04991165(2)\\
\end{tabular}
\end{minipage}
\end{center}
\end{table}

\begin{table}[p]
\begin{center}
\begin{minipage}{12cm}
\caption{Per-diagram contributions to $c_3$.
\label{c3table}}
\begin{tabular}{cr@{}lr@{}lr@{}l}
\multicolumn{1}{c}{Diagram}&
\multicolumn{2}{c}{$c_{3,0}$} &
\multicolumn{2}{c}{$c_{3,1}$} &
\multicolumn{2}{c}{$c_{3,2}$} \\
\tableline \hline
7	&-1&/216                  &0&.00744542215...  	 &-0&.0029334803... \\
8	&0&	                  &0&		         &-0&.0002217811...\\
9       &0&                       &0&.01131461(2)        &-0&.01213765(2)\\
10      &0&                       &0&                    &-0&.0001663358...\\
11      &0&                       &0&                    &-0&.0002217811...\\
12      &0&                       &0&                    &-0&.00016133(1)\\
13      &0&                       &0&                    &-0&.00011990(1)\\
14      &0&                       &0&                    &-0&.00164058(1)\\
15      &0&.0094117909(2)         &-0&.00313726(1)       & 0&.00306459(1)\\
16      &0&                       &0&                    &-0&.00011644(1)\\
17      &0&                       &0&                    &-0&.00008704(1)\\
18      &0&                       &0&                    & 0&.00128600(6)\\
19      &0&                       &0&                    &-0&.00008896(2)\\
20      &0&                       &0&                    &-0&.00004025(1)\\
21      &0&                       &0&                    &0& \\
22-36   &1&/72                    &-0&.03170827(1)       & 0&.01911231(5)\\
\hline
Sum     &0&.0186710502(2)         &-0&.01608550(3)       & 0&.00552737(9)
\end{tabular}
\end{minipage}
\end{center}
\end{table}


\end{document}